\documentclass[12pt]{amsart}
\usepackage{amsmath,amssymb,amsthm,amscd
}
\newcommand\commentout[1]{}
\parskip \medskipamount

\newcommand\ad{\operatorname{ad}}
\newcommand\boldn{\boldsymbol{n}}
\newcommand\calB{{\mathcal B}}
\newcommand\calBbar{\bar{\mathcal B}}
\newcommand\calL{{\mathcal L}}
\newcommand\calLbar{\bar{\mathcal L}}
\newcommand\calM{{\mathcal M}}
\newcommand\calMbar{\bar{\mathcal M}}
\newcommand\calP{{\mathcal P}}
\newcommand\calQ{{\mathcal Q}}
\newcommand\calV{{\mathcal V}}
\newcommand\Complex{{\mathbb C}}
\newcommand\der{\partial}
\newcommand\dmkp{{\mathrm{dmKP}}}

\newcommand\Flag{\mathit{Flag}}
\newcommand\Integer{{\mathbb Z}}

\ifx\mathit\undefined
\newcommand\mathit{} 
\fi

\newcommand\ord{\mathop{\operatorname{ord}}\nolimits}
\newcommand\Res{\mathop{\operatorname{Res}}\nolimits}
\newcommand\SGM{SGM}
\newcommand\tbar{\bar t}

\newcommand\toda{{\mathrm{Toda}}}
\newcommand\ubar{\bar u}
\newcommand\vbar{\bar v}

\newtheorem{thm}{Theorem}[section]
\newtheorem{lem}[thm]{Lemma}
\newtheorem{prop}[thm]{Proposition}

\theoremstyle{definition}

\theoremstyle{remark}

\newtheorem{rem}[thm]{Remark}

\numberwithin{equation}{section}

\newcommand\propref[1]{Proposition~\ref{#1}}
\newcommand\secref[1]{Section~\ref{#1}}

\newcommand\lemref[1]{Lemma~\ref{#1}}

\newcommand\remref[1]{Remark~\ref{#1}}

\begin{document}

\title[Modified KP hierarchy and dispersionless limit]
{A note on the modified KP hierarchy\\
and its (yet another) dispersionless limit}


\author[Takashi TAKEBE]
{Takashi TAKEBE}

\address{
Department of Mathematics,
Faculty of Sciences,
Ochanomizu University,
Otsuka, Bunkyo-ku,
Tokyo, 112-8610, JAPAN
}%

\date{November, 2001}
\ifx\ClassWarning\undefined
\maketitle 
\fi

\begin{abstract}
 The modified KP hierarchies of Kashiwara and Miwa is formulated in Lax
 formalism by Dickey. Their solutions are parametrised by flag
 varieties. Its dispersionless limit is considered.
\end{abstract}

\ifx\ClassWarning\undefined\else
\maketitle
\fi

\setcounter{section}{-1}
\section{Introduction}
\label{sec:introduction}

The {\em modified KP hierarchy} (mKP hierarchy for short) is a system of
non-linear differential equations satisfied by the $\tau$ function
$\tau(s;t) = \langle s | \exp H(t) g |s \rangle$ introduced in early
80's by \cite{kas-miw:81}, \cite{jim-miw:83}. Several Lax
representations of this system have been proposed (cf., for example,
\cite{kuper}, \cite{m-p-z:97} and references in \cite{dic:99}). In this
paper we take a representation in terms of differential-difference
equations by Dickey \cite{dic:99} in a generalised form, which is
suitable for connecting the system to the Toda lattice hierarchy.

As is well-known, the solution space of the KP hierarchy is identified
with the Sato Grassmann manifold (cf.\ \cite{sat:81}, \cite{sat-sat:82},
\cite{sat-nou:84}). The Sato Grassmann manifold consists of subspaces of
an infinite dimensional linear space. We shall see that the solution
space of the mKP hierarchy is the flag varieties consisting of sequences
of subspaces of this linear space.

The dispersionless version of the mKP hierarchy is obtained by the
same procedure as that for the KP and the Toda lattice hierarchies in
\cite{tak-tak:91}, \cite{tak-tak:92}, \cite{tak-tak:93},
\cite{tak-tak:96} and \cite{tak-tak:95}. We will discuss several
features of the dispersionless mKP hierarchies, including the
Riemann-Hilbert type construction (or ``twistor construction'') of
solutions. We shall also see that any solution of the dispersionless
Toda lattice hierarchy is automatically a solution of the modified KP
hierarchies via change of variables.

Starting from representation of the mKP hierarchy different from ours,
various types of dispersionless mKP hierarchy have been obtained. See,
for example, \cite{kuper}, \cite{cha-tu}. Ours is similar to the system
in \cite{kup:92}, but, contrary to Kupershmidt's system, we take the
continuous limit of the lattice.

The author was lead to this subject by a question on Virasoro
constraints posed by Anton Zabrodin, although a satisfactory answer to
his question has not yet been found. The construction presented in
this paper might produce a certain type of Virasoro constraints as in
the case of the dispersionless KP hierarchy. See \cite{tak-tak:92}.

This paper is organised as follows: in the first section we review the
modified KP hierarchies with a slight generalisation and give the
description of the solution space. The dispersionless version of the mKP
hierarchies is introduced in \secref{sec:dmKP}. The third section is
devoted to the Riemann-Hilbert type construction of solutions of the
dispersionless mKP hierarchies. The relation of the dispersionless Toda
lattice hierarchy and the dispersionless mKP hierarchies is discussed in
\secref{sec:dtoda->dmkp}.

\subsection*{Acknowledgements}
The author is grateful to Professor Anton Zabrodin and Professor
Kanehisa Takasaki for stimulating discussions and comments. This work is
partly supported by the Grant-in-Aid for Scientific Research
(No. 13740005), Japan Society for the Promotion of Science.

\section{Modified KP hierarchy: Review and Generalisation}
\label{sec:mKP}

Fix a non-empty subset $\boldn$ of $\Integer$. We number the elements of
$\boldn$ in the increasing order as $\{n_0, n_1, \dots,n_N\}$ ($n_0 <
n_1 < \dots < n_N$) when the cardinality of $\boldn$ is finite, as
$\{n_0, n_1, \dots\}$ ($n_s < n_{s+1}$) when $\boldn$ is an infinite set
and has the minimum number, and otherwise as $\{n_s\}_{s\in\Integer}$
($n_s < n_{s+1}$). We denote $m_s=n_{s+1}-n_s$ when both $s$ and $s+1$
are in $S$, where $S$ is the index set of $\boldn$. 

Let us introduce one continuous variable $x$, a series of continuous
independent variables $t= (t_1,t_2,\dots)$ and a discrete independent
variable $s \in S$. We denote $\der/\der x$ by $\der$. The dependent
variables $u_n(t;s)$ and $p_k(t;s)$ are encapsulated in the Lax
operator,
\begin{equation}
    L(s)=L(t;s) 
    := \der + u_2(t;s) \der^{-1} + u_3(t;s) \der^{-2} + \cdots
    = \sum_{n=1}^\infty u_n(t;s) \der^{1-n},   
\label{def:L}
\end{equation}
with $u_0=1$, $u_1 = 0$, and in the auxiliary differential operator,
\begin{equation}
    P(s)=P(t;s) 
    := \der^{m_s} + p_{m_s-1}(t;s) \der^{m_s-1} + \cdots 
    + p_1(t;s) \der + p_0(t;s),
\label{def:P}
\end{equation}
for $s$. (We do not write the $x$-dependence explicitly by the reason we
shall mention in the next paragraph.) We always assume $s, s+1 \in S$
whenever the operator $P(s)$ or the number $m_s$ appear.

The Lax representation of the {\em $\boldn$-modified KP hierarchy}
($\boldn$-mKP hierarchy for short) is the system
\begin{align}
    \frac{\der L(s)}{\der t_n} &= [B_n(s), L(s)], \qquad
    B_n(s) := (L(s)^n)_{\geq 0},
\label{mkp:lax:L}
\\
    L(s+1) P(s) &= P(s) L(s),
\label{mkp:LP=PL}
\\
    (\der_{t_n} - B_n(s+1)) P(s) &= P(s) (\der_{t_n} - B_n(s)),
\label{mkp:BP=PB}
\end{align}
where $(\cdot)_{\geq 0}$ denotes the usual projection of a
micro-differential operator to the differential operator part. The first
equation \eqref{mkp:lax:L} is nothing but the Lax representation of the
KP hierarchy and the remaining two equations, \eqref{mkp:LP=PL} and
\eqref{mkp:BP=PB}, determine the consistency of $L(s)$. The Lax
equations \eqref{mkp:lax:L} and \eqref{mkp:BP=PB} for $n=1$ means that
$t_1$ and $x$ appear always in a combination $t_1+x$, so we shall omit
$x$ unless it is necessary.

This formulation was introduced by Dickey \cite{dic:99} when $\boldn =
\Integer$. When $\boldn$ consists of a single number, e.g., $\boldn=
\{0\}$, the $\boldn$-mKP hierarchy is the KP hierarchy. If
$\boldn=\{n_s\}_{s\in S} \subset \boldn'=\{n'_s\}_{s\in S'}$, a solution
of the $\boldn'$-mKP hierarchy gives a solution of the $\boldn$-mKP
hierarchy: let $\iota: S \to S'$ be an injection such that
$n'_{\iota(s)} = n_s$. Then $(L(\iota(s)), \tilde P(s))_{s\in S}$ is a
solution of the $\boldn$-mKP hierarchy for any solution
$(L(s),P(s))_{s\in S'}$ of the $\boldn'$-mKP hierarchy. Here $\tilde
P(s)$ is a differential operator defined by $\tilde P(s) =
P(\iota(s+1)-1) \cdots P(\iota(s)+1) \, P(\iota(s))$.

Hereafter we fix $\boldn$ and omit it unless mentioned otherwise.

\begin{rem}
\label{rem:Lax}
 The Lax equations \eqref{mkp:lax:L} are redundant since we can recover
 all of them from the equation \eqref{mkp:lax:L} for $s=0$,
 \eqref{mkp:LP=PL} and \eqref{mkp:BP=PB}. See Proposition 2.2 of
 \cite{dic:99}.
\end{rem}

\begin{rem}
\label{rem:log(P)}
 If we rewrite the commutation relation \eqref{mkp:LP=PL} as
\begin{equation}
    (L(s+1) - L(s)) P(s) = [P(s), L(s)],
\label{mkp:lax:log(P)}
\end{equation}
 we can formally interpret it as a Lax type equation, ``$L(s+1)-L(s) =
 [\log P(s), L(s)]$''.
\end{rem}

The mKP hierarchy is the consistency condition of the following system
of linear equations for $w(t;s;\lambda)$:
\begin{equation}
 \begin{split}
    &L(s) w(t;s;\lambda) = \lambda w(t;s;\lambda),
\qquad
    \frac{\der}{\der t_n} w(t;s;\lambda) = B_n(s) w(t;s;\lambda),
\\
    &P(s) w(t;s;\lambda) = w(t;s+1;\lambda).
 \end{split}
\label{mkp:lin}
\end{equation}
By the argument well-known in the theory of the KP hierarchy, we have a
monic $0$-th order micro-differential operator, $W(s) = W(t;s) = 1 +
w_1(t;s)\der^{-1} + w_2(t;s)\der^{-2} + \cdots$, which satisfies
\begin{equation}
    L(S) = W(s) \der W(s)^{-1}, \qquad
    \frac{\der W(s)}{\der t_n} = - (L(s)^n)_{<0} W(s),
\label{mkp:sato}
\end{equation}
where $(\cdot)_{<0}$ is the projection to the negative order
part. Moreover, the equations \eqref{mkp:LP=PL} and \eqref{mkp:BP=PB}
assures that we can adjust $W(s)$ so that $P(s)$ is expressed as
\begin{equation}
    P(s) = W(s+1) \der^{m_s} W(s)^{-1}.
\label{P=WdW}
\end{equation}
Hence, the linear problem \eqref{mkp:lin} has a solution (the {\em wave
function}) of the form,
\begin{equation}
 \begin{split}
    w(t;s;\lambda) &:= W(s) \der^{n_s} e^{\xi(t;\lambda)}
    = \hat w(t;s;\lambda) \lambda^{n_s} e^{\xi(t;\lambda)},
\\
    \hat w(t;s;\lambda) &:=
    1 + w_1(t;s)\lambda^{-1} + w_2(t;s)\lambda^{-2} + \cdots,
 \end{split}
\label{def:wave-func}
\end{equation}
where $\xi(t;\lambda)= \sum_{n=1}^\infty t_n\lambda^n$. The adjoint wave
function is defined by
\begin{equation}
 \begin{split}
    w^\ast(t;s;\lambda) 
    &:= (W(s)^\ast)^{-1} (-\der)^{n_s} e^{-\xi(t;\lambda)}
    = \hat w^\ast(t;s;\lambda) \lambda^{n_s} e^{-\xi(t;\lambda)},
\\
    \hat w^\ast(t;s;\lambda) &:=
    1 + w_1^\ast(t;s)\lambda^{-1} + w_2^\ast(t;s)\lambda^{-2} + \cdots,
 \end{split}
\label{def:adj-wave-func}
\end{equation}
where $W(s)^\ast = 1 + (-\der)\circ w_1^\ast(t;s) + (-\der)^2\circ
w_2^\ast(t;s) + \cdots$ is the formal adjoint operator of $W(s)$.

In terms of the wave operator $W(s)$, the mKP hierarchy is rewritten as 
\begin{equation}
    \frac{\der W(s)}{\der t_n} = - (W(s)\der^nW(s)^{-1})_{<0} W(s),
    \qquad
    (W(s+1) \der^{m_s} W(s)^{-1})_{<0} = 0.
\label{mkp:W}
\end{equation}
This system is expressed as the bilinear residue identity, which
corresponds to (1.5.1) of \cite{djkm} for the KP hierarchy: two
functions $w(t;s;\lambda)$ and $w^\ast(t;s;\lambda)$ of the form
\eqref{def:wave-func} and \eqref{def:adj-wave-func} are a wave function
and its adjoint if and only if they satisfy
\begin{equation}
    \oint_{\lambda=\infty}
    w(t;s;\lambda)\, w^\ast(t';s';\lambda)\, d\lambda = 0,
\label{bil-res}
\end{equation}
for any $t$, $t'$ and $s \geqq s'$. The proof is essentially the same as
that in \S1.5 of \cite{djkm} or the proof of Proposition 2.6 and 2.7 of
\cite{dic:99}, which rewrites \eqref{mkp:W} by Lemma 1.1 of \cite{djkm}
cited as Lemma 2.5 in \cite{dic:99}.

The Orlov-Schulman operator (cf.\ \cite{orl-sch:86}, \cite{tak-tak:95})
is defined by
\begin{equation}
 \begin{split}
    M(t;s) &:= 
    W(t;s) 
    \left( \sum_{n=1}^{\infty} n t_n \der^{n-1} + x + n_s \der^{-1}\right)
    W(t;s)^{-1}
\\
    &= \sum_{n=1}^{\infty} n t_n L^{n-1} + x + n_s L^{-1}
    +  \sum_{n=1}^{\infty} v_n(t;s) L^{-n-1}
\end{split}
\label{def:M}
\end{equation}
and satisfies
\begin{equation}
    \frac{\der M}{\der t_n} = [B_n, M],
\qquad
    M(s+1) P(s) = P(s) M(s),
\qquad
    [L(s), M(s)] = 1.
\label{mkp:M}
\end{equation}
The wave function defined by \eqref{def:wave-func} satisfies
\begin{equation}
     M(s) w(t;s;\lambda) = \frac{\der}{\der x}w(t;s;\lambda),
\label{mkp:lin:M}
\end{equation}
The Orlov-Schulman operator plays an important role when we consider the
dispersionless limit. As in \cite{orl-sch:86}, we can describe the
symmetries of the mKP hierarchy by using $L$ and $M$ but we do not go
to this direction here.

There exists the $\tau$ function, $\tau(t;s)$, which satisfies
\begin{equation}
   \hat w(t;s;\lambda) =
   \frac{\tau(t-[\lambda^{-1}];s)}{\tau(t;s)}.\qquad
   \hat w^\ast(t;s;\lambda) =
   \frac{\tau(t+[\lambda^{-1}];s)}{\tau(t;s)}.
\label{w=tau/tau}
\end{equation}
This is the direct consequence of the theory of the KP hierarchy. See
p.~269 of \cite{sat-sat:82} or \S1.6 of \cite{djkm}. Note that we have a
gauge freedom,
\begin{equation}
    \tau(t;s) \mapsto \tau(t;s) e^{\Lambda(s)},
\label{mkp:tau:gauge}
\end{equation}
where $\Lambda(s)$ is an arbitrary function of $s$ and independent of
$t_n$ ($n=1,2,\dots$). We can write down the bilinear equations
characterising the $\tau$ function by substituting the expressions
\eqref{w=tau/tau} in \eqref{bil-res}, which we leave to the reader. The
representation theoretical description of the $\tau$ function of the
modified KP hierarchy goes back to \cite{kas-miw:81} and
\cite{jim-miw:83}.

As is discussed in \cite{uen-tak:84}, a solution of the Toda lattice
hierarchy gives a series of solutions of the KP hierarchy parametrised
by a discrete variable $s$, if one fixes the half set of the continuous
independent variables. In fact, this series is a solution of the
$\boldn$-mKP hierarchy defined above where $\boldn = \Integer$. The
differential operator $P(t;s)$ is the operator $\der - b_0(s)$ at the
end of \S1.2 of \cite{uen-tak:84}.

While each solution of the KP hierarchy corresponds to a point on the
Sato Grassmann manifold, the solution of the mKP hierarchy corresponds
to a point on an infinite dimensional flag variety. More precise
statement is as follows. Let $\calV$ be an infinite dimensional linear
space and $V^\emptyset$ is its subspace defined by $\calV =
\bigoplus_{\nu\in\Integer} \Complex e_\nu$, $\calV^{(0)} =
\bigoplus_{\nu \geqq 0} \Complex e_\nu$.  (Actually we have to take the
completion of $\calV$, but details are omitted.) The {\em Sato Grassmann
manifold} of charge $n$, $\SGM^{(n)}$, is defined by
\begin{equation}
    \SGM^{(n)} =
    \{ U \subset \calV \mid
    \text{ index of\ }U \to \calV/\calV^{(0)} \text{ is\ }n
    \}.
\label{def:SGM}
\end{equation}
The solution space of the KP hierarchy is $\SGM^{(0)}$ as is shown in
\cite{sat:81}, \cite{sat-sat:82} or \cite{sat-nou:84}.

\begin{prop}
\label{prop:flag}
 Each solution of the $\boldn$-mKP hierarchy for $W(t;s)$,
 \eqref{mkp:W}, is parametrised by the flag variety,
\begin{equation}
    \Flag_{\boldn}
    :=
    \{ (U_s)_{s\in S} \mid
    U_s \in \SGM^{(n_s)}, \ 
    U_s \subset U_{s+1}\}.
\end{equation}
\end{prop}

The proof is the same as that of Corollary 3.4 of \cite{take:91}.

\section{Dispersionless limit of modified KP hierarchy}
\label{sec:dmKP}

The dispersionless version of the mKP hierarchy is obtained by the same
procedure as that for the KP and the Toda lattice hierarchies in
\cite{tak-tak:91}, \cite{tak-tak:92}, \cite{tak-tak:93},
\cite{tak-tak:96} and \cite{tak-tak:95}. Since the discrete parameter
$s$ of the mKP hierarchy turns into a continuous parameter, we need to
start from the $\boldn$-mKP hierarchy with constant $m_s = n_{s+1}-n_s$,
which is the order of the differential operator $P(t;s)$. We denote this
number by $N$ and assume $\boldn = N \Integer = \{n_s =
Ns\}_{s\in\Integer}$.

First, let us introduce a parameter $\hbar$ and rewrite the equations
\eqref{mkp:lax:L}, \eqref{mkp:LP=PL} (or \eqref{mkp:lax:log(P)}),
\eqref{mkp:BP=PB} and \eqref{mkp:M}, rescaling the independent variables
$t$ and $s$, as follows:
\begin{align}
    &\hbar \frac{\der L}{\der t_n} = [B_n, L], \qquad
    \hbar\frac{\der M}{\der t_n} = [B_n, M], \qquad
    B_n := (L^n)_{\geq 0},
\label{mkph:lax}
\\
    &[L(s), M(s)] = \hbar,
\label{mkph:ccr}
\\
    &(L(s+\hbar)-L(s)) P(s) = [P(s), L(s)],
\label{mkph:LP=PL}
\\
    &(M(s+\hbar) - M(s)) P(s) = [P(s), M(s)],
\label{mkph:MP=PM}
\\
    &- (B_n(s+\hbar) - B_n(s)) P(s) = [P(s), \hbar \der_{t_n} - B_n(s)],
\label{mkph:BP=PB}
\end{align}
where the Lax operator $L(t;s)$, the Orlov-Schulman operator $M(t;s)$
and the differential operator $P(t;s)$ have the following form:
\begin{align}
    L(t;s) 
    &= \sum_{n=1}^\infty u_n(t;s) (\hbar\der)^{1-n},
\label{def:hL}
\\
    M(t;s) 
    &= \sum_{n=1}^{\infty} n t_n L^{n-1} + x + N s L^{-1}
    +  \sum_{n=1}^{\infty} v_n(t;s) L^{-n-1}
\label{def:hM}
\\
    P(t;s) 
    &= 
    (\hbar\der)^{N} + p_{N-1}(t;s)(\hbar\der)^{N-1} + \cdots
    + p_0(t;s).
\label{def:hP}
\end{align}
The leading terms of \eqref{mkph:lax}, \eqref{mkph:ccr},
\eqref{mkph:LP=PL}, \eqref{mkph:MP=PM} and \eqref{mkph:BP=PB} with
respect to the order ($\ord \hbar = -1$, $\ord \der/\der x = \ord
\der/\der t_n = 1$) give the {\em dispersionless $N$-modified KP
hierarchy}. The operators $L(t;s)$, $M(t;s)$ and $P(t;s)$ are replaced
by formal series in variable $k$,
\begin{align}
    \calL(t;s) 
    &= \sum_{n=1}^\infty u_n(t;s) k^{1-n}
     = k + u_2(t;s) k^{-1} + \cdots,
\label{def:calL}
\\
    \calM(t;s) 
    &= \sum_{n=1}^{\infty} n t_n \calL^{n-1} + x + Ns \calL^{-1}
    +  \sum_{i=1}^{\infty} v_{i}(t;s) \calL^{-i-1},
\label{def:calM}
\\
    \calP(t;s) &= k^N + p_{N-1}(t;s) k^{N-1} + \cdots + p_0(t;s).
\label{def:calP}
\end{align}
(We set $v_0(t;s) = Ns$.) The system of differential-difference
equations \eqref{mkph:lax} etc.\ become the differential equations with
respect to $t$ and $s$:
\begin{align}
    &\frac{\der \calL}{\der t_n} = \{\calB_n, \calL\}, \qquad
    \frac{\der \calM}{\der t_n} = \{\calB_n, \calM\}, \qquad
    \calB_n := (\calL^n)_{\geq 0},
\label{dmkp:lax}
\\
    &\{\calL(s), \calM(s)\} = 1,
\label{dmkp:ccr}
\\
    &\frac{\der \calL}{\der s} \calP(s) = \{\calP(s), \calL(s)\},
    \qquad
    \frac{\der \calM}{\der s} \calP(s) = \{\calP(s), \calM(s)\},
\label{dmkp:[L,M,P]}
\\
    &- \frac{\der \calB_n}{\der s} \calP(s) 
    = - \frac{\der \calP}{\der t_n} 
    - \{ \calP(s), \calB_n(s) \},
\label{dmkp:BP=PB}
\end{align}
where $(\cdot)_{\geq 0}$ is the projection of power series in $k$ to the
polynomial part and $\{\cdot,\cdot\}$ is the same Poisson bracket as
the Poisson bracket for the dispersionless KP hierarchy:
\begin{equation}
    \{f(k,x), g(k,x)\} 
    = 
    \frac{\der f}{\der k}\frac{\der g}{\der x}
    -
    \frac{\der f}{\der x}\frac{\der g}{\der k}.
\label{def:poisson}
\end{equation}
The three equations, \eqref{dmkp:[L,M,P]} and \eqref{dmkp:BP=PB} can be
formally interpreted as follows (cf.\ \remref{rem:log(P)}):
\begin{gather}
    \frac{\der \calL}{\der s} = \{\log \calP(s), \calL(s)\}, \qquad
    \frac{\der \calM}{\der s} = \{\log \calP(s), \calM(s)\},
\label{dmkp:[L,M,logP]}
\\
    \frac{\der}{\der t_n} \log \calP - \frac{\der \calB_n}{\der s}
    +
    \{\log \calP, \calB_n\} = 0.
\label{dmkp:[B,logP]}
\end{gather}

We can also define the dispersionless $N$-mKP hierarchy as the system
for $\calL$ and $\calP$ only. Namely, it is enough to take the equations
\eqref{dmkp:lax}, \eqref{dmkp:[L,M,P]} and \eqref{dmkp:BP=PB} for
$\calL$ and $\calP$. It is possible to introduce the dressing operator
$\exp \ad\varphi$ as in \cite{tak-tak:95}. ($\ad f(g) = \{f,g\}$.) The
series $\varphi(t;s) = \sum_{n=1}^\infty \varphi_n(t) k^{-n}$ satisfies
\begin{equation}
 \begin{split}
    \nabla_{t_n,\varphi} \varphi 
    &= \calB_n - e^{\ad\varphi} k^n
     = -(e^{\ad\varphi} (k^n))_{<0},
\\
    \nabla_{s,\varphi} \varphi 
    &= \log \calP -e^{\ad\varphi} \log k^N,
 \end{split}
\label{dressing}
\end{equation}
where
$
    \nabla_{u, \psi} \varphi 
    = \sum_{m=0}^\infty 
      \frac1{(m+1)!} (\ad\psi)^m
       \frac{\der \varphi}{\der u},
$
for series $\psi$ and $\varphi$ and a variable $u$. The symbol
$(\cdot)_{< 0}$ denotes the projection to the negative power part as
usual. Equations \eqref{dmkp:lax} (for $\calL$), \eqref{dmkp:[L,M,P]}
(for $\calL$) and \eqref{dmkp:BP=PB} are the compatibility conditions of
\eqref{dressing}.  To define the Orlov-Schulman function $\calM$ we have
only to put
\begin{equation}
    \calM = 
    e^{\ad\varphi} 
    \left(\sum_{n=1}^\infty n t_n k^{n-1}+ x + Ns k^{-1}\right).
\label{calM}
\end{equation}

\begin{lem}
\label{lem:d(vn)}
The coefficient $v_n$ in \eqref{def:calM} satisfies
\begin{equation}
    \frac{\der v_n}{\der t_m} 
    = \Res_{k=\infty} \calL^n d\calB_m,
\qquad
    \frac{\der v_n}{\der s} 
    = \Res_{k=\infty} \calL^n d\log\calP.
\label{d(vn)}
\end{equation}
\end{lem}
The first equation is Proposition 4 of \cite{tak-tak:92} and the second
equation is proved in the same way. Using these relations, we can show
that $\calB_n$ and $\calP$ are expanded with respect to $\calL$ as
follows: 
\begin{equation}
    \calB_n =
    \calL^n + 
    \sum_{m=1}^\infty 
    \frac{-1}{m} \frac{\der v_m}{\der t_n} \calL^{-m},
\qquad
    \calP =
    \calL^N \exp\left(
    \sum_{m=1}^\infty 
    \frac{-1}{m} \frac{\der v_m}{\der s} \calL^{-m}
    \right).
\label{calB,calP:calL-expansion}
\end{equation}
The first equation is (4.7) of \cite{tak-tak:92}, while the second is
proved by applying the similar argument to $\log (\calP/\calL^N)$
instead of $\calB_n$.

It follows from \lemref{lem:d(vn)} and Proposition 6 of
\cite{tak-tak:92} (a consequence of \eqref{d(vn)};
$
    \der v_n/\der t_m = \der v_m/\der t_n
$
) that the system for a function $\phi(t;s)$
\begin{equation}
    \frac{\der \phi}{\der t_n} = \Res_{k=\infty} \calL^n d\log\calP,
\label{def:phi}
\end{equation}
is consistent. The $s$-dependence of $\phi$ is not fixed by these
equations and we may add any function of $s$ independent of $t_n$ to
$\phi(t;s)$. 

The $\tau$ function for the dispersionless mKP hierarchy is defined by
\begin{equation}
    d\log \tau_{\dmkp} (t;s) 
    = \sum_{n=1}^\infty v_n(t;s) dt_n + \phi(t;s) ds,
\label{def:dmkp-tau}
\end{equation}
where the exterior differentiation $d$ is taken with respect to the
variables $t_n$ ($n=1,2,\dots$) and $s$. Because of the additive
ambiguity of $\phi(t;s)$ mentioned above, $\log\tau_\dmkp(t;s)$ has the
gauge freedom:
\begin{equation}
    \log\tau_\dmkp(t;s) \mapsto 
    \log\tau_\dmkp(t;s) + \Lambda(s),
\label{dmkp:tau:gauge}
\end{equation}
for an arbitrary function $\Lambda(s)$ of $s$. This corresponds to the
gauge freedom of the $\tau$ function of the mKP hierarchy,
\eqref{mkp:tau:gauge}. 

\section{Twistor construction of solutions of the dmKP hierarchy} 
\label{sec:twistor}

Any solution $(\calL, \calM)$ of the dispersionless KP hierarchy is
obtained from a pair of functions $(f(k,x),g(k,x))$ (``twistor data'')
with the canonical Poisson relation $\{f,g\} = 1$ by requiring
\begin{equation}
    (f(\calL, \calM))_{<0} = 0, \qquad
    (g(\calL, \calM))_{<0} = 0,
\label{twistor:dkp}
\end{equation}
which was shown in \cite{tak-tak:92} and \cite{tak-tak:95}. We call this
method ``the twistor construction''\footnote{For recent developments of
twistor construction of solutions of the dispersionless KP equations,
see \cite{dun-tod:01} and \cite{dun:01}.}  or ``the Riemann-Hilbert type
construction'' of solutions.

This theorem implies that any solution of the dispersionless mKP
hierarchy $(\calL(s), \calM(s), \calP(s))$ should be associated to
twistor data $(f(k,x,s),g(k,x,s))$ depending on the variable
$s$, satisfying
\begin{equation}
    (f(\calL(s), \calM(s),s))_{<0} = 0, \qquad
    (g(\calL(s), \calM(s),s))_{<0} = 0.
\label{twistor:dmkp:1}
\end{equation}
The dependence of the series $f(k,x,s)$ and $g(k,x,s)$ on $s$ is given
by the following proposition. Hereafter we use the notations like
\begin{equation*}
 \begin{split}
    \frac{\der f}{\der s} (\calL(s),\calM(s),s) &:=
    \left. \frac{\der f}{\der s}(k,x,s) \right|_{k=\calL(s),x=\calM(s)},
\\
    \frac{\der}{\der s} f (\calL(s),\calM(s),s) &:=
    \frac{\der}{\der s}\bigl(f (\calL(s),\calM(s),s) 
                   \text{ as a function of $s$}\bigr).
 \end{split}
\end{equation*}
We define
\begin{equation}
    \hat \calL(s) = f(\calL(s), \calM(s), s), \qquad
    \hat \calM(s) = g(\calL(s), \calM(s), s),
\label{def:Lhat,Mhat}
\end{equation}
which consists of positive powers of $k$ if \eqref{twistor:dmkp:1} is
satisfied.

\begin{prop}
\label{prop:twistor:dmkp}
 If the series $\calL(s)$, $\calM(s)$ and the polynomial $\calP(s)$ of
 the form \eqref{def:calL}, \eqref{def:calM} and \eqref{def:calP}
 satisfy \eqref{twistor:dmkp:1} and 
\begin{equation}
 \begin{split}
    &\left(
    \calP(s) \frac{\der}{\der s} \hat\calL(s)
    + \{\calP(s), \hat\calL(s) \}
    -
    \calP(s) \frac{\der f}{\der s}(\calL(s),\calM(s),s)
    \right)_{<N-1} = 0,
\\
    &\left(
    \calP(s) \frac{\der}{\der s} \hat\calM(s)
    + \{\calP(s), \hat\calM(s) \}
    -
    \calP(s) \frac{\der g}{\der s}(\calL(s),\calM(s),s)
    \right)_{<N-1} = 0,
 \end{split}
\label{twistor:dmkp:2}
\end{equation}
 then $(\calL(s),\calM(s),\calP(s))$ is a solution of the dispersionless
 mKP hierarchy. Here $(\cdot)_{<N-1}$ is the projection to the Laurent
 series in $k$ with powers less than $N-1$.
\end{prop}
Note that the contents in the parentheses of \eqref{twistor:dmkp:2}
vanish for any function $f$ and $g$ if $(\calL,\calM,\calP)$ is a
solution of the dispersionless mKP hierarchy.

The conditions \eqref{twistor:dmkp:2} are not so neat as
\eqref{twistor:dmkp:1}, but when $N=1$, they reduce to the
following due to \eqref{twistor:dmkp:1}: 
\begin{equation}
    \left(
    \calP(s) \frac{\der f}{\der s}(\calL(s),\calM(s),s)
    \right)_{<0} = 0,
\qquad
    \left(
    \calP(s) \frac{\der g}{\der s}(\calL(s),\calM(s),s)
    \right)_{<0} = 0.
\label{twistor:dmkp:3}
\end{equation}

\begin{proof}[Proof of \propref{prop:twistor:dmkp}]
 The conditions \eqref{twistor:dmkp:1} assure that $(\calL(s),
 \calM(s))$ are solutions of the dispersionless KP hierarchy,
 \eqref{dmkp:lax} and \eqref{dmkp:ccr} by virtue of Proposition 7 of
 \cite{tak-tak:92}. Therefore we have only to check the equations
 \eqref{dmkp:[L,M,P]} and \eqref{dmkp:BP=PB}.

 Differentiating \eqref{def:Lhat,Mhat} with respect to $s$, we have
\begin{equation}
    \begin{pmatrix}
    \dfrac{\der\hat\calL}{\der s} - \dfrac{\der f}{\der s}(\calL,\calM,s)
    \\
    \dfrac{\der\hat\calM}{\der s} - \dfrac{\der g}{\der s}(\calL,\calM,s)
    \end{pmatrix}
    =
    \begin{pmatrix}
    \dfrac{\der f(\calL,\calM,s)}{\der\calL} &
    \dfrac{\der f(\calL,\calM,s)}{\der\calM} \\
    \dfrac{\der g(\calL,\calM,s)}{\der\calL} &
    \dfrac{\der g(\calL,\calM,s)}{\der\calM} 
    \end{pmatrix}
    \begin{pmatrix}
    \dfrac{\der \calL}{\der s} \\
    \dfrac{\der \calM}{\der s}
    \end{pmatrix}.
\label{d(Lhat,Mhat)/ds}
\end{equation}
 According to the equation (6.5) of \cite{tak-tak:92}, the $2\times 2$
 matrix in the right hand side is decomposed as
\begin{equation}
\begin{split}
    \begin{pmatrix}
    \dfrac{\der f(\calL,\calM,s)}{\der\calL} &
    \dfrac{\der f(\calL,\calM,s)}{\der\calM} \\
    \dfrac{\der g(\calL,\calM,s)}{\der\calL} &
    \dfrac{\der g(\calL,\calM,s)}{\der\calM} 
    \end{pmatrix}
    &=
    \begin{pmatrix}
    \dfrac{\der \hat\calL}{\der k} &
    \dfrac{\der \hat\calL}{\der x} \\
    \dfrac{\der \hat\calM}{\der k} &
    \dfrac{\der \hat\calM}{\der x}
    \end{pmatrix}
    \begin{pmatrix}
    \dfrac{\der \calL}{\der k} &
    \dfrac{\der \calL}{\der x} \\
    \dfrac{\der \calM}{\der k} &
    \dfrac{\der \calM}{\der x}
    \end{pmatrix}^{-1}
\\
    &=
    \begin{pmatrix}
     \dfrac{\der \hat\calM}{\der x} &
    -\dfrac{\der \hat\calL}{\der x} \\
    -\dfrac{\der \hat\calM}{\der k} &
     \dfrac{\der \hat\calL}{\der k}
    \end{pmatrix}^{-1}
    \begin{pmatrix}
     \dfrac{\der \calM}{\der x} &
    -\dfrac{\der \calL}{\der x} \\
    -\dfrac{\der \calM}{\der k} &
     \dfrac{\der \calL}{\der k}
    \end{pmatrix},
 \end{split}
\label{d(f,g)/d(L,M)}
\end{equation}
 where we used the canonical Poisson relations $\{\calL(s),\calM(s)\} =
 \{\hat\calL(s), \hat\calM(s)\} = 1$ (cf.\ (6.8) of \cite{tak-tak:92}). 
 Substituting \eqref{d(f,g)/d(L,M)} into \eqref{d(Lhat,Mhat)/ds} and
 multiplying $\calP(s)$, we have
\begin{equation}
    \begin{pmatrix}
     \dfrac{\der \hat\calM}{\der x} &
    -\dfrac{\der \hat\calL}{\der x} \\
    -\dfrac{\der \hat\calM}{\der k} &
     \dfrac{\der \hat\calL}{\der k}
    \end{pmatrix}
    \begin{pmatrix}
      \calP(s)\dfrac{\der\hat\calL}{\der s} 
    - \calP(s)\dfrac{\der f}{\der s}(\calL,\calM,s)
    \\
      \calP(s)\dfrac{\der\hat\calM}{\der s}
    - \calP(s)\dfrac{\der g}{\der s}(\calL,\calM,s)
    \end{pmatrix}
    = \calP(s)
    \begin{pmatrix}
     \dfrac{\der \calM}{\der x} \dfrac{\der \calL}{\der s} 
    -\dfrac{\der \calL}{\der x} \dfrac{\der \calM}{\der s} \\
    -\dfrac{\der \calM}{\der k} \dfrac{\der \calL}{\der s}
    +\dfrac{\der \calL}{\der k} \dfrac{\der \calM}{\der s}
    \end{pmatrix}.
\label{twistor:positive}
\end{equation}
 The upper component of the vector in the right hand side is, by the
 computation similar to (6.12), (6.13) of \cite{tak-tak:92}, rewritten
 as follows:
\begin{equation}
 \begin{split}
    &\frac{\der \calM}{\der x} \dfrac{\der \calL}{\der s} 
    -\frac{\der \calL}{\der x} \dfrac{\der \calM}{\der s}
\\
    =&
    \left(
      \frac{\der \calL}{\der s} 
    + \sum_{n=1}^\infty \frac{1}{-n} \frac{\der v_{n}}{\der x}
      \frac{\der \calL^{-n}}{\der s}
    \right) 
    -
    \left(
      N \frac{\der \log \calL}{\der x}
    + \sum_{n=1}^\infty \frac{1}{-n} \frac{\der v_{n}}{\der s}
      \frac{\der \calL^{-n}}{\der x}
    \right) 
\\ 
    =&
    \frac{\der}{\der s}\left(
      \calL
    + \sum_{n=1}^\infty 
      \frac{1}{-n} \frac{\der v_{n}}{\der x} \calL^{-n}
    \right) 
    -
    \frac{\der}{\der x}\left(
      N  \log\calL
    + \sum_{n=1}^\infty 
      \frac{1}{-n} \frac{\der v_{n}}{\der s} \calL^{-n}
    \right).
 \end{split}
\label{twistor:positive:rhs1:temp}
\end{equation}
 Thus we have
\begin{equation}
     \frac{\der \calM}{\der x} \dfrac{\der \calL}{\der s} 
    -\frac{\der \calL}{\der x} \dfrac{\der \calM}{\der s}
    = -\frac{\der}{\der x}\log\calQ,
\label{twistor:positive:rhs1}
\end{equation}
 due to (4.9) and (5.6) of \cite{tak-tak:92}, where
\begin{equation}
    \calQ = \calL^N \exp\left(
    \sum_{n=1}^\infty
    \frac{1}{-n} \frac{\der v_{n}}{\der s} \calL^{-n}\right).
\label{def:calQ}
\end{equation}
 (Reminding \eqref{calB,calP:calL-expansion}, one would expect that
 $\calQ$ should be $\calP$, which will turn out to be true.) The lower
 component of the vector in the right hand side of
 \eqref{twistor:positive} is, by the same computation,
\begin{equation}
     \frac{\der \calM}{\der k} \frac{\der \calL}{\der s} 
    -\frac{\der \calL}{\der k} \frac{\der \calM}{\der s}
    = - \frac{\der}{\der k}\log\calQ.
\label{twistor:positive:rhs2}
\end{equation}
 Thus, substituting \eqref{twistor:positive:rhs1} and
 \eqref{twistor:positive:rhs2} into \eqref{twistor:positive} and adding
 the equation
\begin{equation*}
    \begin{pmatrix}
     \dfrac{\der \hat\calM}{\der x} &
    -\dfrac{\der \hat\calL}{\der x} \\
    -\dfrac{\der \hat\calM}{\der k} &
     \dfrac{\der \hat\calL}{\der k}
    \end{pmatrix}
    \begin{pmatrix}
    \{\log\calP(s), \hat\calL(s)\} \\ \{\log\calP(s), \hat\calM(s)\}
    \end{pmatrix}
    =
    \begin{pmatrix}
    \dfrac{\der}{\der x}\log\calP \\ 
    \dfrac{\der}{\der k}\log\calP
    \end{pmatrix},
\end{equation*}
 we obtain
\begin{equation}
 \begin{split}
    &
    \begin{pmatrix}
     \dfrac{\der \hat\calM}{\der x} &
    -\dfrac{\der \hat\calL}{\der x} \\
    -\dfrac{\der \hat\calM}{\der k} &
     \dfrac{\der \hat\calL}{\der k}
    \end{pmatrix}
    \begin{pmatrix}
      \calP(s)\dfrac{\der\hat\calL}{\der s} 
    + \{\calP(s), \hat\calL(s)\}
    - \calP(s)\dfrac{\der f}{\der s}(\calL,\calM,s)
    \\
      \calP(s)\dfrac{\der\hat\calM}{\der s}
    + \{\calP(s), \hat\calM(s)\}
    - \calP(s)\dfrac{\der g}{\der s}(\calL,\calM,s)
    \end{pmatrix}
\\
    =& \calP(s)
    \begin{pmatrix}
    -\dfrac{\der}{\der x}\log(\calQ/\calP) \\
    -\dfrac{\der}{\der k}\log(\calQ/\calP)
    \end{pmatrix}.
 \end{split}
\label{twistor:Q=P}
\end{equation}
 This is the point where the condition \eqref{twistor:dmkp:2} plays the
 role. According to this condition, the left hand side does not contain
 $k^m$ with $m < N-1$. Because of the normalisation \eqref{def:calP} of
 $\calP$ and the definition \eqref{def:calQ} of $\calQ$, the second
 component of the right hand side of \eqref{twistor:Q=P} is a Laurent
 series of order not greater than $N-1$, which proves
$
    \frac{\der}{\der k}\log(\calQ/\calP) = 0.
$
 Hence $\calP=\calQ \times $(constant independent of $k$). The
 normalisations of $\calP$ and $\calQ$ implies $\calP=\calQ$. Thus the
 equations \eqref{twistor:positive:rhs1} and
 \eqref{twistor:positive:rhs2} take the form
\begin{equation}
    \begin{pmatrix}
     \dfrac{\der \hat\calM}{\der x} & -\dfrac{\der \hat\calL}{\der x} \\
     \dfrac{\der \hat\calM}{\der k} & -\dfrac{\der \hat\calL}{\der k}
    \end{pmatrix}
    \begin{pmatrix}
    \dfrac{\der \calL}{\der s} \\  \dfrac{\der \calM}{\der s}
    \end{pmatrix}
    = 
    \begin{pmatrix}
    - \dfrac{\der}{\der x}\log\calP \\
    - \dfrac{\der}{\der k}\log\calP
    \end{pmatrix},
\end{equation}
 from which follows \eqref{dmkp:[L,M,P]} by the
 unimodularity of the matrix in the left hand side.

 Lastly we prove \eqref{dmkp:BP=PB}. By decomposing $\calL^n$ as
 $\calB_n + (\calL)_{<0}$, this equation is rewritten as
\begin{equation}
    \frac{\der }{\der t_n} \log(\calP/\calL^N)
    +
    \frac{\der}{\der s} (\calL^n)_{<0}
    -
    \{\log(\calP/\calL^N), (\calL^n)_{<0}\} = 0.
\label{BP=PB:<0}
\end{equation}
 Note that the relations \eqref{calB,calP:calL-expansion} hold for our
 $\calL$ and $\calP$, since the first equation is the result of the
 theory of the dispersionless KP hierarchy only and the second equation
 is the result of \eqref{def:calQ} and $\calP=\calQ$. Hence we may
 substitute them into \eqref{BP=PB:<0}. The rest of the proof is a
 straightforward computation using the Lax equation \eqref{dmkp:lax}.
\end{proof}

\section{Relation with the dispersionless Toda lattice hierarchy}
\label{sec:dtoda->dmkp}

In this section we show that any solution of the dispersionless Toda
lattice hierarchy provides a solution of the dispersionless mKP
hierarchy. 

Before discussing this relation, recall that the relation among the
(dispersionful) mKP hierarchies mentioned in \S1: if $\boldn \subset
\boldn' \subset \Integer$, then a solution of the $\boldn'$-mKP
hierarchy gives a solution of the $\boldn'$-mKP hierarchy. The
dispersionless version of this relation is almost trivial: suppose $N =
m N'$ for a positive integer $m$. If $(\calL(s),\calP(s))$ is a solution
of the dispersionless $N$-mKP hierarchy, setting $s'= ms$ and
$\tilde\calP(s') = \calP(s)^m$, we have
\begin{equation*}
    \frac{\der \calL}{\der s'} = \{\log\tilde\calP(s'), \calL\},
    \qquad
     \frac{\der}{\der t_n} \log\tilde\calP(s')
    - \frac{\der\calB_n}{\der s'} 
    +\{\log\tilde\calP(s'), \calB_n\} = 0,
\end{equation*}
which means that $(\calL(s'), \tilde\calP(s'))$ is a solution of the
dispersionless $N'$-mKP hierarchy. Hence to show that the dispersionless
$N$-mKP hierarchy is embedded in the dispersionless Toda lattice
hierarchy, it suffices to show the case $N=1$, which we always assume in
this section.

The dispersionless Toda lattice hierarchy is defined as follows:
It is a system of differential equations with two sets of infinite
variables $t = (t_1, t_2, \ldots)$ and $\tbar = (\tbar_1, \tbar_2,
\ldots)$ and one more variable $s$. The Lax representation is:
\begin{equation}
 \begin{split}
    \frac{\der \calL}{\der t_n} 
    = \{ \calB_n, \calL \}_\toda, 
    \quad&
    \frac{\der \calL}{\der \tbar_n} 
    = \{ \calBbar_n, \calL \}_\toda,
\\
    \frac{\der \calLbar}{\der t_n} 
    = \{ \calB_n, \calLbar \}_\toda,
    \quad&
    \frac{\der \calLbar}{\der \tbar_n} 
    = \{ \calBbar_n, \calLbar \}_\toda,
    \quad n = 1,2,\ldots,
 \end{split}
\label{dtoda:lax}
\end{equation}
where $\calL$ and $\calLbar$ are Laurent series
\begin{equation}
    \calL         = p
           + \sum_{n=0}^\infty u_{n+1}     (t,\tbar,s) p^{-n},
\qquad
    \calLbar^{-1} =  \ubar_0(t, \tbar, s) p^{-1} 
           + \sum_{n=0}^\infty \ubar_{n+1} (t,\tbar,s) p^n,
\label{dtoda:L}
\end{equation}
of a variable $p$, and $\calB_n$ and $\calBbar_n$ are given by $\calB_n
= (\calL^n)_{\ge 0}$, $\calBbar_n = (\calLbar^{-n})_{< 0}$.  Here
$(\cdot)_{\ge 0}$ and $(\cdot)_{< 0}$ denote the projection of power
series to a positive power part and a negative power part
respectively. The Poisson bracket $\{\cdot,\cdot\}_\toda$ is defined by
\begin{equation}
    \{ A(p,s), B(p,s) \}_\toda
    = p\frac{\der A(p,s)}{\der p} \frac{\der B(p,s)}{\der s}
     -p\frac{\der A(p,s)}{\der s} \frac{\der B(p,s)}{\der p}.
\label{dtoda:poisson} 
\end{equation}
The Orlov-Schulman functions $\calM$ and $\calMbar$ are of the form
\begin{equation}
 \begin{split}
    \calM &= \sum_{n=1}^\infty n t_n \calL^n + s
           + \sum_{n=1}^\infty v_n(t,\tbar,s) \calL^{-n},
\\
    \calMbar &= - \sum_{n=1}^\infty n \tbar_n \calLbar^{-n} + s
                + \sum_{n=1}^\infty \vbar_n(t,\tbar,s) \calLbar^n.
 \end{split}
\label{dtoda:M}
\end{equation}
They satisfy the Lax equations \eqref{dtoda:lax} with $\calL$ and
$\calLbar$ replaced by $\calM$ and $\calMbar$, and the canonical Poisson
relations $\{\calL,\calM\}_\toda = \calL$, $\{\calLbar,\calMbar\}_\toda
= \calLbar$. 

Proposition 2.8.1 of \cite{tak-tak:95} asserts that, if $(\calL,
\calLbar)$ is a solution of the dispersionless Toda hierarchy
\eqref{dtoda:lax}, then $\calL(t,\tbar,s)$ is a solution of the
dispersionless KP hierarchy \eqref{dmkp:lax} when we identify
$\calB_1=p+u_1(t,\tbar,s)$ with $k$:
\begin{equation}
    \left.\frac{\der \calL}{\der t_n}\right|_{k:\mathrm{ fixed}} =
    \{ \calB_n, \calL \}.
\label{dtoda->dkp} 
\end{equation}
Here $\{\cdot,\cdot\}$ is the Poisson bracket for the dispersionless KP
hierarchy, \eqref{def:poisson}. Note that the projection onto a
polynomial in $p$ is equal to the projection onto a polynomial in $k$,
since $k = \calB_1 = p + u_1(t,\tbar,s)$.  Thus $\calB_n =
(\calL^n)_{\geq 0}$ in the sense of \eqref{dmkp:lax} as well as in the
sense of \eqref{dtoda:lax}.

In fact, it is a direct computation to check that the above solution of
the dispersionless KP hierarchy $\calL(s)$ together with
$\calP=p=k-u_1(t,\tbar,s)$ gives a solution of the dispersionless mKP
hierarchy \eqref{dmkp:lax}, \eqref{dmkp:[L,M,P]} and \eqref{dmkp:BP=PB}.

As is shown in \cite{tak-tak:91}, if there is a pair of functions
$(f_\toda(p,s),g_\toda(p,s))$ in $(p,s)$ which satisfies $\{f,g\}_\toda
= f$, and the series $\calL$, $\calLbar$, $\calM$, $\calMbar$ of the
form \eqref{dtoda:L} and \eqref{dtoda:M} satisfy
\begin{equation}
    f_{\toda}(\calL_\toda,\calM_\toda) = \bar\calL_\toda, \qquad
    g_{\toda}(\calL_\toda,\calM_\toda) = \bar\calM_\toda,
\label{twistor:dtoda}
\end{equation}
then $(\calL_\toda,\calLbar_\toda,\calM_\toda,\calMbar_\toda)$ gives a
solution of the dispersionless Toda lattice hierarchy (``twistor
construction of solutions'').

\begin{prop}
\label{prop:twistor:dtoda->dmkp}
 Let $(f_\toda(p,s), g_\toda(p,s))$ be the above twistor data for the
 dispersionless Toda lattice hierarchy and $(\calL_\toda,
 \bar\calL_\toda, \calM_\toda,\bar\calM_\toda)$ be the corresponding
 solution. Then $(f_{\dmkp}(k,x,s), g_{\dmkp}(k,x,s))$ defined by
\begin{equation}
 \begin{split}
    f_{\dmkp}(k,x,s)&:= f_{\toda}(k,kx),
\\
    g_{\dmkp}(k,x,s)&:= (g_{\toda}(k,kx) - s) f_{\toda}(k,kx)^{-1},
 \end{split}
\label{twistor:dtoda->dmkp}
\end{equation}
 satisfies the canonical Poisson relation. 

 Let us define the triplet $(\calL_\dmkp, \calM_\dmkp, \calP_\dmkp)$ by
\begin{equation}
    \calL_\dmkp = \calL_\toda, \qquad
    \calM_\dmkp=\calM_\toda \calL_\toda^{-1}, \qquad
    \calP_\dmkp = k - u_0(t;0;s),
\label{L,M,P:dtoda->dmkp}
\end{equation} 
 with all $p$ in $\calL_\toda$, $\calM_\toda$ replaced by $k -
 u_0(t;0;s)$. Then it gives a solution of the dispersionless mKP
 hierarchy corresponding to $(f_\dmkp,g_\dmkp)$ by
 \propref{prop:twistor:dmkp}.
\end{prop}

The proof of the canonical Poisson relation is nothing more than an
elementary computation. The fact that the triplet $(\calL_\dmkp,
\calM_\dmkp, \calP_\dmkp)$ satisfies \eqref{twistor:dmkp:1} and
\eqref{twistor:dmkp:3} is trivial by the definition of $(\calL_\dmkp,
\calM_\dmkp, \calP_\dmkp)$ and $(f_{\dmkp}, g_{\dmkp})$.

For example, the twistor data of the dispersionless Toda lattice
hierarchy applied to the two-dimensional string theory in \cite{taka:95}
and to the interface dynamics in \cite{m-w-z:00} is
\begin{equation}
    f_\toda(p,s) = ps^{-1}, \qquad g_\toda(p,s) = s.
\label{data:2dstring}
\end{equation}
\propref{prop:twistor:dtoda->dmkp} says that the solution of the
dispersionless mKP hierarchy obtained from this solution by the change
of variables solves the Riemann-Hilbert type problem
\eqref{twistor:dmkp:1} and \eqref{twistor:dmkp:2} for 
\begin{equation}
    f_\dmkp(k,x,s) = x^{-1}, \qquad g_\dmkp(k,x,s) = (kx-s) x.
\label{data:2dstring:dmkp}
\end{equation}

\section{Concluding remarks}
\label{sec:conclusion}

We have shown that the basic properties of the dispersionless KP and
Toda hierarchies hold also for the dispersionless mKP hierarchies. The
$w_{\infty+1}$-symmetries in the dispersionless KP and Toda hierarchies
should also be found easily in the modified case, which should be
inherited from the $W_{\infty+1}$-symmetries of the mKP hierarchies (or
the Toda lattice hierarchy), though we did not discuss them in this
paper.

Apparently small but maybe essential discrepancy with the KP/Toda case
arises when we consider solutions. The last example
\eqref{data:2dstring:dmkp} in \secref{sec:dtoda->dmkp} shows that the
twistor data for the dispersionless mKP hierarchies can be rather
complicated. (Recall that the analysis of the Virasoro constraint was
possible in \cite{tak-tak:92} because the twistor data are at most
linear in $\calM$.)

Problems like construction of the $(f,g)$-pair for the $M$-reduction
($\calL^M = $ a polynomial of $k$, namely, $f(k,x,s) = k^M$) and
analysis of the Virasoro constraints for such solutions still remain
open. One of the reason why this is not so trivial is the normalisation
$v_0(t;s) = N s$ in \eqref{def:calM}. To attack such problems, it might
be necessary to refine \propref{prop:twistor:dmkp}.


\end{document}